\documentclass[prl,twocolumn,floatfix,showpacs]{revtex4-1}
\usepackage{graphicx}
\usepackage{dcolumn}
\usepackage{bm}
\usepackage{color}
\usepackage{natbib}
\usepackage{array}
\usepackage{subfigure}
\usepackage{amssymb}
\usepackage{amsbsy}
\usepackage{natbib}
\usepackage{epstopdf}
\usepackage{placeins}
\usepackage{footnote}
\usepackage{citesort}
\usepackage{hyperref}
\usepackage{url}
\begin{document}

\setlength{\parskip}{0pt}
\setlength{\floatsep}{4pt}
\setlength{\textfloatsep}{4pt}
\setlength{\abovecaptionskip}{0pt}
\setlength{\belowcaptionskip}{0pt}

\title{Thermal boundary layer profiles in turbulent Rayleigh-B\'enard convection in a cylindrical sample}
\author{Richard J.A.M. Stevens$^1$}
\author{Quan Zhou$^{2}$}
\author{Siegfried Grossmann$^3$}
\author{Roberto Verzicco$^{1,4}$}
\author{Ke-Qing Xia$^5$}
\author{Detlef Lohse$^1$}
\affiliation{$^1$Physics of Fluids Group, Department of Science and Technology and J.M. Burgers Center for Fluid Dynamics, University of Twente, P.O Box 217, 7500 AE Enschede, The Netherlands\\
$^2$ Shanghai Institute of Applied Mathematics and Mechanics, Shanghai University, Shanghai 200072, China\\
$^3$ Fachbereich Physik, Philipps-Universit\"{a}t Marburg, D-35032 Marburg, Germany\\
$^4$ Dept. of Mech. Eng., Universita' di Roma "Tor Vergata",Via del Politecnico 1, 00133, Roma, Italy\\
$^5$ Department of Physics, The Chinese University of Hong Kong, Shatin, Hong Kong, China}
\date{\today}

\begin{abstract}
We numerically investigate the structures of the near-plate temperature profiles close to the bottom and top plates of turbulent Rayleigh-B\'{e}nard flow in a cylindrical sample at Rayleigh numbers $Ra=10^8$ to $Ra=2\times10^{12}$ and Prandtl numbers $Pr=6.4$ and $Pr=0.7$ with the dynamical frame method [Q. Zhou and K.-Q. Xia, {\it Phys. Rev. Lett.} {\bf 104}, 104301 (2010)] thus extending previous results for quasi-2-dimensional systems to 3D systems for the first time. The dynamical frame method shows that the measured temperature profiles in the spatially and temporally local frame are much closer to the temperature profile of a laminar, zero-pressure gradient boundary layer according to Pohlhausen than in the fixed reference frame. The deviation between the measured profiles in the dynamical reference frame and the laminar profiles increases with decreasing $Pr$, where the thermal BL is more exposed to the bulk fluctuations due to the thinner kinetic BL, and increasing $Ra$, where more plumes are passing the measurement location.
\end{abstract}

\maketitle

One of the classical systems to study heat transport phenomena is turbulent Rayleigh-B\'enard (RB) convection. In this system a layer of fluid is heated from below and cooled from above \cite{ahl09,loh10}. The dynamics and the global features of the system are strongly influenced by the properties of the flow in the boundary layers (BLs). Almost all theories that describe the heat transport in turbulent RB convection, from the early marginal stability theory \cite{mal54} to the Shraiman $\&$ Siggia (SS) model \cite{shr90,sig94} and to the Grossmann $\&$ Lohse (GL) theory \cite{gro-all}, are essentially BL theories. Therefore, it is a key issue to fully understand BL profiles close to the horizontal plates.

The key question in RB convection is how do the heat transport, expressed by the Nusselt number $Nu$, and the turbulence intensity, expressed by the Reynolds number $Re$, depend on the control parameters of the system? For given aspect ratio $\Gamma\equiv D/L$ ($D$ is the sample diameter and $L$ its height) and given geometry, the control parameters are the Rayleigh number $Ra=\beta g\Delta L^3 /(\kappa \nu)$ and the Prandtl number $Pr=\nu/\kappa$. Here $\beta$ is the thermal expansion coefficient, $g$ the gravitational acceleration, $\Delta$ the temperature difference between the plates, and $\nu$ and $\kappa$ are the kinematic viscosity and thermal diffusivity, respectively. The GL theory has achieved good success in predicting $Nu (Ra,Pr)$ and $Re (Ra,Pr)$ \cite{fun05,xia02,ahl09}.

Recently, the GL theory was successfully extended to the very large $Ra$ number regime \cite{gro11} (the so called ultimate range), in order to explain the experimentally observed multiple scaling regimes of the heat transfer \cite{he11} and to the rotating case to predict the heat transfer enhancement due to rotation \cite{ste10b}. As the GL theory heavily builds on the assumption that the BL thickness of not yet turbulent BLs scales inversely proportional to the square root of the $Re$ number according to Prandtl's 1904 theory \cite{pra04}, the degree of validity of Prandtl-Blasius BL \cite{bla08} flow needs to be tested. Indeed, recent experiments \cite{sun08} have shown that in non-rotating RB the BLs scaling behaves as in laminar flows. In addition, a comparison of the mean bulk temperature calculated using the Prandtl-Blasius theory with that measured in both liquid and gaseous non-Oberbeck-Boussinesq RB convection shows very good agreement \cite{ahl06,ahl07,ahl08}. Furthermore, the kinematic BL thickness evaluated by solving the laminar Prandtl-Blasius BL equations was found to agree well with that obtained in the direct numerical simulation (DNS) \cite{shi10}.
However, a comparison between experimental velocity \cite{pui07b} and numerical temperature \cite{shi09} profiles obtained in the fixed reference frame (fixed with respect to the horizontal plates) with the respective classical profiles shows significant deviations.

\begin{figure*}
  \centering
  \subfigure{\includegraphics[width=0.95\textwidth]{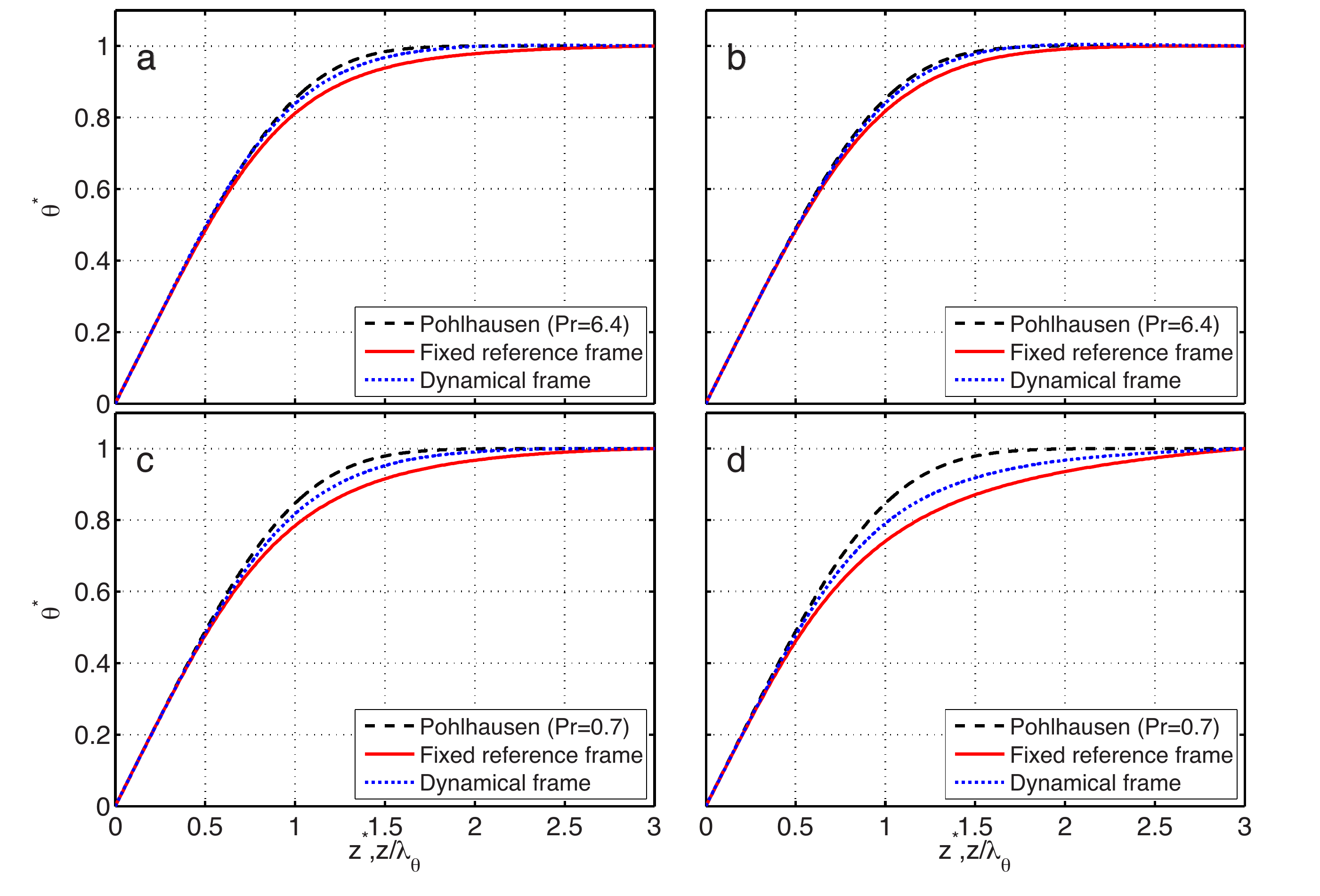}}
  \caption{Comparison between temperature profiles obtained at the cell center axis near the bottom and top plate: dynamical $\theta^{*}(z^{*})$ (blue short dashed lines), fixed reference frame $\theta(z)$ (red solid lines), and the temperature profile of a laminar, zero-pressure gradient boundary layer according to Pohlhausen (black long dashed lines) for a (a) $Pr=6.4$, $Ra=10^8$, and $\Gamma=1$, (b) $Pr=6.4$, $Ra=10^8$, and $\Gamma=1/2$, (c) $Pr=0.7$, $Ra=10^8$, and $\Gamma=1/2$ (d) $Pr=0.7$, $Ra=2\times10^{12}$, and $\Gamma=1/2$.}
  \label{figure1}
\end{figure*}

\begin{figure}
  \centering
  \subfigure{\includegraphics[width=0.45\textwidth]{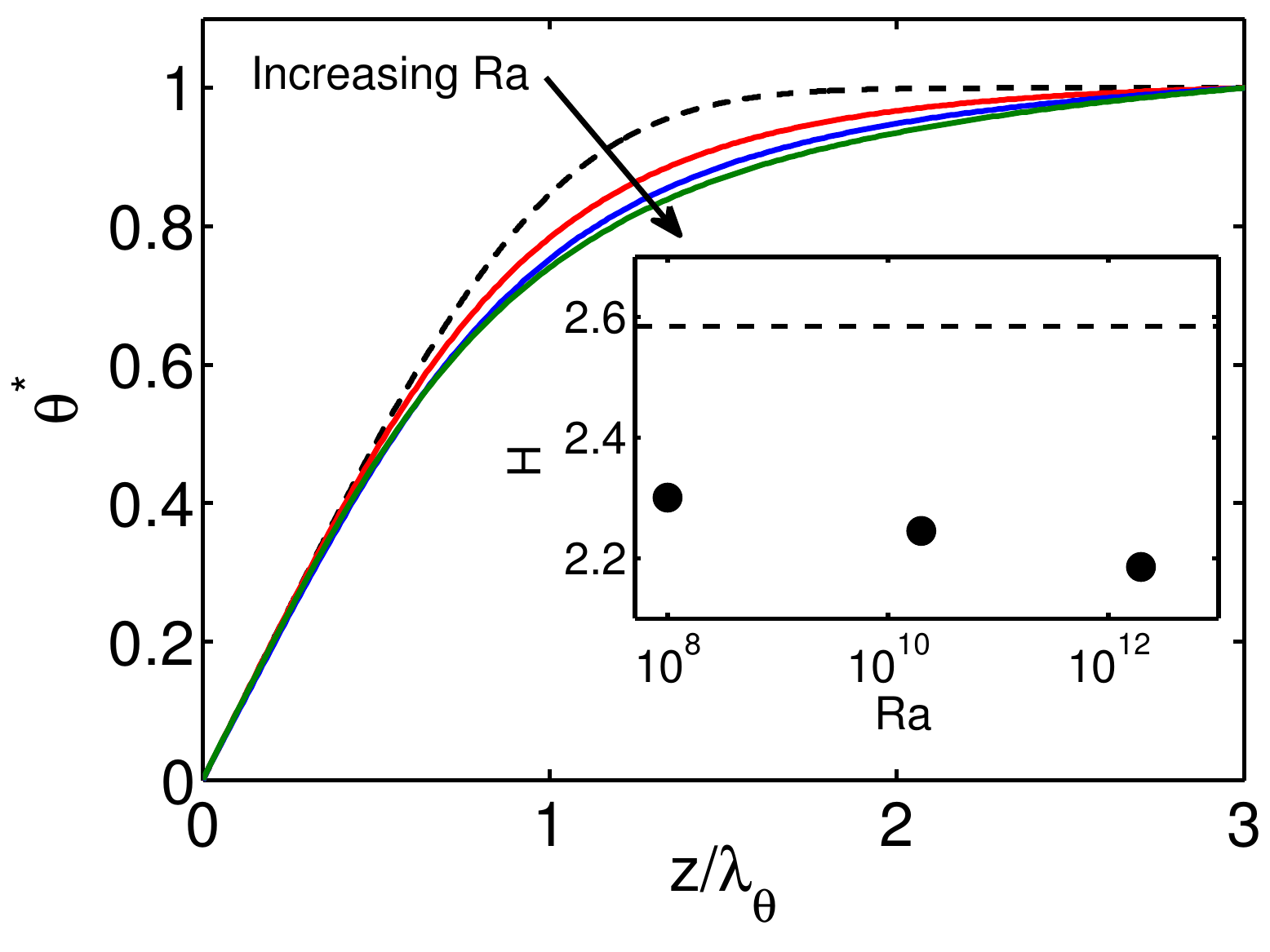}}
  \subfigure{\includegraphics[width=0.45\textwidth]{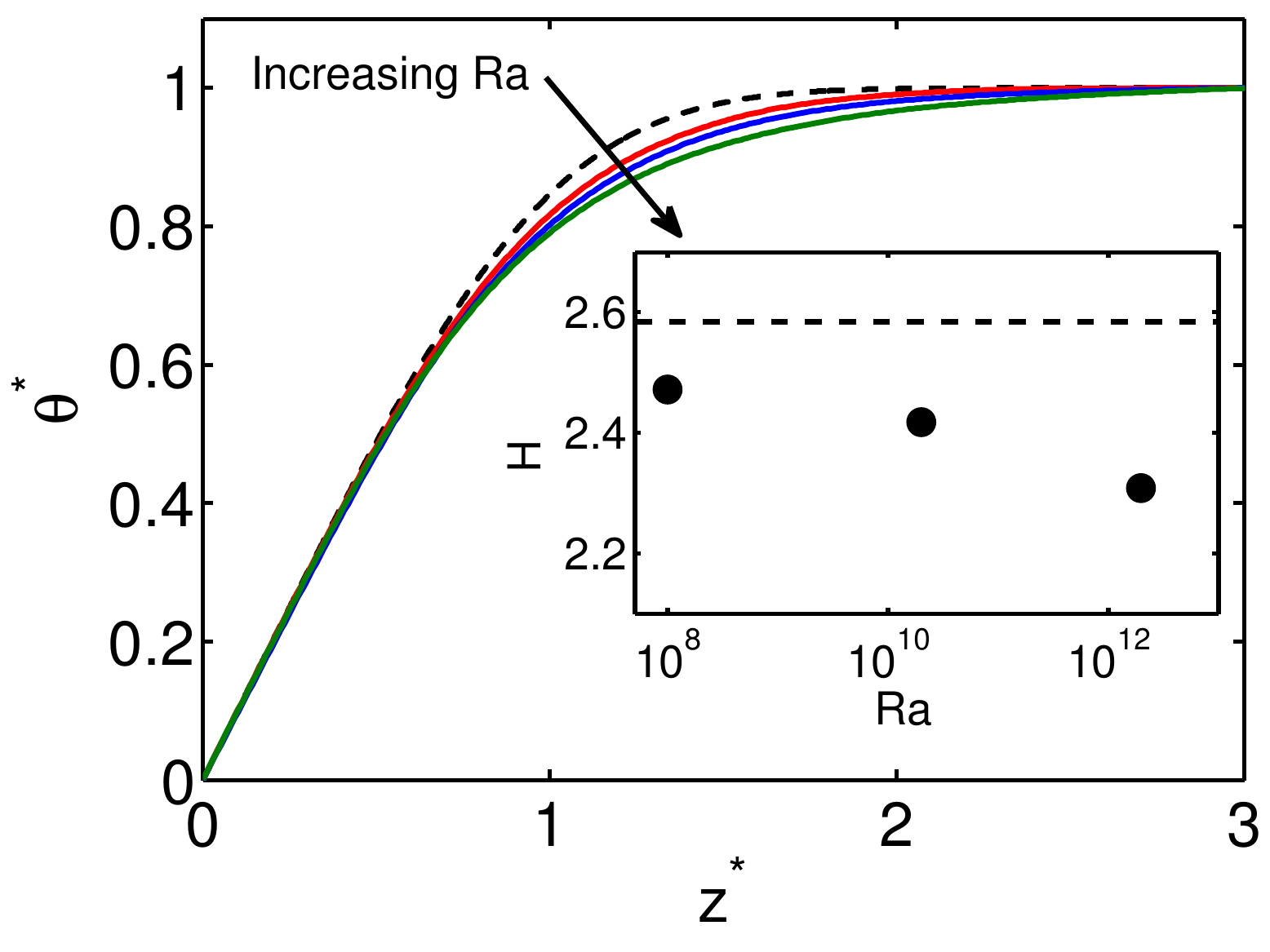}}
  \caption{Comparison between temperature profiles obtained at the cell center near the bottom and top plate in the fixed reference frame (a) and the dynamical reference frame (b) for $Pr=0.7$ in an aspect ratio $\Gamma=1/2$ sample at $Ra=10^8$ (red), $Ra=2\times10^{10}$ (blue), $Ra=2\times10^{12}$ (dark green) and the temperature profile of a laminar, zero-pressure gradient boundary layer with $Pr=0.7$ according to Pohlhausen (dashed line). Note that the difference between this theoretical profile and the numerical data increases with increasing $Ra$ as indicated by the direction of the arrow. The insets in figure (a) and (b) compares the shape factor for the temperature profile of a laminar, zero-pressure gradient boundary layer with $Pr=0.7$ according to Pohlhausen (horizontal dashed line) with the shape factor of the temperature profiles obtained in the fixed reference frame (panel a) and the dynamical reference frame (panel b). Note that the shape factor in the dynamical reference frame is closer to the theoretical prediction than in the fixed reference frame.}
  \label{figure2}
\end{figure}

Considerable progress on this issue was recently achieved by Zhou and Xia \cite{zho10} who experimentally studied the velocity BL for water ($Pr=4.3$) with Particle Image Velocimetry (PIV). They argued that such deviations should be attributed to intermittent emissions of thermal plumes from the BLs and the corresponding temporal dynamics of the BLs \cite{sug09, zho10}. This led to the study of the BL structures in dynamical reference frames, which fluctuate with the instantaneous BL thicknesses \cite{zho10}. Zhou et al.\ \cite{zho10} used PIV data to show that the velocity profiles obtained in the dynamical reference frame are very close to the laminar Prandtl-Blasius profile and numerical data \cite{zho10b,zho11} to show that a good agreement with the temperature profile of a laminar, zero-pressure gradient boundary layer according to Pohlhausen \cite{zho10b,zho11}, which we will call the "Pohlhausen" profile \cite{poh21} in the remainder of this paper is obtained . In addition, they showed that most of the time the rescaled instantaneous velocity and temperature profiles are also in agreement with the theoretical prediction. A dynamical BL rescaling method has thus been established, which extends the time-independent Prandtl-Blasius-Pohlhausen BL theory to the time-dependent case, in the sense that it holds locally at every instant in the frame that fluctuates with the local instantaneous BL thickness.

However, this dynamical rescaling method has up to now only been applied to experimental and numerical data obtained in (quasi-) 2D RB convection, where the orientation of the large scale circulation (LSC) is fixed. In addition, the experimental result is limited to $Ra = 3 \times 10^{11}$. In this paper we will apply the dynamical rescaling method to previously obtained data from fully 3D DNS performed in a cylindrical sample. The questions we will address here are: Does the dynamical rescaling method also work for this 3D geometry, where more complex flow structures can form? Does the method work for the highest $Ra$ number simulations, i.e. $Ra = 2 \times10^{12}$ we have available? And is a similar agreement obtained also for lower $Pr$?

For the analysis presented here we use data from our previous simulations, i.e. the simulations of Stevens et al.\ \cite{ste10,ste10d} for $Pr=0.7$ in an aspect ratio $\Gamma=1/2$ sample for $Ra=2\times10^6$ to $Ra = 2\times 10^{12}$, and the simulations of Stevens et al.\ \cite{ste10c} for $Ra=10^8$, $Pr=0.7$ and $Pr=6.4$, in $\Gamma=1$ and $\Gamma=1/2$ samples. All these simulations have been performed with a second order finite difference code, see Verzicco et al.\ \cite{ver96,ver97,ver99} for details. During the simulations mentioned above we stored, besides a limited number of full 3D snapshots, a large number of vertical (and horizontal) snapshots over a period of $30$ to $150$ dimensionless time units when time is scaled by $L/\sqrt{\beta g \Delta L}$. The vertical snapshots can be used to analyze the temperature profile in the BL at the center axis. This is a favorable location as the LSC is always passing the center axis, provided that the off-center (or sloshing) motion of the LSC is not too strong near the plates \cite{xi09,zho09}. In addition, restricting the analysis to the temperature field offers the benefit that the orientation of the LSC just above the plates does not have to be determined, because the temperature profile should be independent of the azimuthal orientation of the LSC in the region around the center axis. This means that no difficult (and arbitrary) routines have to be used to determine the orientation and presence of the LSC. However, a drawback of this method is that we cannot determine whether the LSC is passing the cell center at every time instance. As is shown by Zhou et al.\ \cite{zho10,zho10b,zho11} the agreement between the dynamically rescaled profile and the Prandtl-Blasius-Pohlhausen profile deteriorates when one considers a position where the LSC does not pass. Similar effects can influence our results, however preventing this is very difficult as it would require an arbitrary criterion to determine the presence of the LSC at a certain location.

We determined the mean temperature profile in the dynamical reference frame by defining the instantaneous thermal BL thickness $\delta_{th}(t)$ as the intersection between the temperature gradient at the plate and the temperature of the first extremum value in the temperature profile \cite{zho10b,zho11}. Even then we obtain that the mean bulk temperature obtained far away from the thermal BL is below $\theta^*$. As discussed by Zhou et al.\, \cite{zho10b,zho11} we attribute this to the emissions of thermal plumes. They propose to use the temperature at some position outside of the thermal BL, such as $z/\lambda_{th}=3$ or $z^*_{th}=3$, as the asymptotic value for the BL, where $\lambda_{th}$ is the slope BL thickness obtained in the fixed reference frame and $z^*_{th}(t)\equiv z/\delta_{th}(t)$ is the rescaled vertical distance from the plate. In figure \ref{figure1} we show a direct comparison between the various temperature profiles, rescaled in the described way: the fixed reference frame profiles $\theta(x, z)/\theta(x, z=3\lambda_{th})$, the dynamical local profiles $\theta^*(z_{th}^*)/\theta^*(z_{th}^*=3)$, where $\theta^*$ indicates the temperature in the dynamical reference frame defined as $\theta^*(x,z^*_{th})\equiv\langle \theta(x,z=z^*_{th}\delta_{th}(x,t),t)\rangle$ \cite{zho10,zho10b,zho11}, and the Pohlhausen temperature profile.

Figure \ref{figure1}a and \ref{figure1}b shows that for $Ra=10^8$ and $Pr=6.4$ the mean dynamically-rescaled profiles are very close to the Pohlhausen temperature profile profile for both $\Gamma=1$ and $\Gamma=1/2$ samples. In agreement with the results obtained by Zhou et al.\ \cite{zho10,zho10b,zho11} in (quasi) 2D RB convection we find a significant preference of the dynamical frame based profiles: Around the thermal BL thickness the fixed reference frame based time averaged local profiles $\theta(z)/\theta(z=3\lambda_{th})$ are all much lower than the Pohlhausen profile, while the dynamically rescaled instantaneous local profiles $\theta^*(z_{th}^*)/\theta^*(z_{th}^*=3)$ are much closer to the Pohlhausen profile. We emphasize that we observe this good agreement even though we do not filter out the moments in which the LSC is not passing the cell center. In addition, one has to realize that buoyancy driven instabilities generate a wall normal flow and break down the stratified structure of a laminar flow field in the BLs. Consequently, heat is no longer mainly conducted but also a significant fraction is convected throughout the BLs. This results in a deviation of the mean temperature profiles from the laminar BL predictions as seen in the figure. Another feature worthy of comment is that the results here show that the dynamic scaling method works also for $\Gamma =1/2$ geometry, whereas all previous studies \cite{zho10,zho10b,zho11} only considered a $\Gamma=1$ geometry. That the method still works in a $\Gamma =1/2$ sample is remarkably given that  the laminar BL theory has been developed for parallel flow over an infinite flat plate, whereas here in the aspect ratio $\Gamma=1/2$ sample one can hardly find such regions of parallel flows at the top and bottom plates.

In figure \ref{figure1}c we show the results for $Pr=0.7$ and $Ra=10^8$ in a $\Gamma=1/2$ sample. A comparison with the results for $Pr=6.4$ at the same $Ra$ reveals that  for the lower $Pr$ the dynamically rescaled profile differs more, but still less than in the fixed reference frame, from the Pohlhausen temperature profile. The larger deviation at lower Pr can be explained by the fact that the viscous BL is thicker than the thermal BL for larger Pr and hence the thermal BL in this case is less influenced by the bulk turbulent flow than at low $Pr$. Figure \ref{figure1}d shows that the difference between the dynamically rescaled profile and the Pohlhausen profile increases with increasing $Ra$.
To quantify the deviations from the Pohlhausen profiles we determined the shape factor of the temperature profiles in the BLs close to the bottom and top plate, cf \cite{sch00}
\begin{equation}
H\equiv\delta^d/\delta^m,
\end{equation}
where $\delta^d$ and $\delta^m$ are the local displacement and momentum thicknesses of the profiles, respectively, defined as,
\begin{equation}
\delta^d\equiv\int_0^{\infty} \left[1-\frac{Y(z)}{[Y(z)]_{max}}\right] dz
\end{equation}
and
\begin{equation}
\delta^m\equiv\int_0^{\infty}\ \left[1-\frac{Y(z)}{[Y(z)]_{max}}\right] \left[\frac{Y(z)}{[Y(z)]_{max}}\right] dz.
\end{equation}
Here, $Y=\theta^*$ or $\theta$ is the corresponding local temperature profile in the dynamical or fixed reference frame, and all $z$-integrations are evaluated only over the range from $z=0$ to $z=3$ instead of towards $\infty$. Roughly speaking, the shape factor of a profile in general describes how fast the profile approaches its asymptotic value, i.e. the larger the shape factor is, the faster the profile runs to its asymptotic level. The shape factor of the Pohlhausen temperature profile is $H^{PB}\approx2.59$ for the present $Pr=0.7$. Figure \ref{figure2} shows that the shape factor is closer to the Pohlhausen value in the dynamical reference frame than in the fixed reference frame, which confirms that the profiles obtained in the dynamical reference frame are indeed closer to the Pohlhausen profile. Here we emphasize that the peak in the PDF of the shape factor is even closer to the shape factor expected from the BL theory, e.g. for $Ra=2\times10^{10}$ the peak in the PDF of the shape factor in the dynamical reference frame is at $H\approx 2.50$, whereas the average shape factor in the dynamical reference frame is $H\approx 2.42$. Similar numbers are observed for the other $Ra$ numbers. Furthermore, we see that the shape factor decreases for increasing $Ra$, which confirms that the difference between the dynamically rescaled profile and the Pohlhausen profile indeed increases with increasing $Ra$. We attribute this to the fact that the flow becomes more turbulent at lower $Pr$ and higher $Ra$ and a more turbulent flow leads to the formation of a larger number of plumes that pass the center axis. Because Zhou et al.\ \cite{zho10,zho10b,zho11} have shown that passing plumes lead to a lower shape factor, it is not unexpected that the difference between the dynamically rescaled profile and the Pohlhausen prediction increases with $Ra$. In fact both for lower $Pr$ and higher $Ra$ the velocity field  becomes more turbulent and advects the temperature differently than described by purely laminar theory which can leads to the larger deviations that are observed. We note, that even for the most turbulent case we considered here, i.e. $Ra=2 \times 10^{12}$ and $Pr=0.7$, we cannot identify any characteristic signature indicating that the BLs have become turbulent. In particular we do not observe logarithmic profiles, which would look significantly different, i.e. very steep near the wall and much flatter with extensions of order $L$ away from the wall. Both these properties we do not observe in the simulations. This is in agreement with recent experimental \cite{he11} and theoretical \cite{gro11} results, which show that the BLs become turbulent around $10^{13} \leq Ra \leq 5\times 10^{14}$. However, we cannot exclude that the observed deviations in the BL profiles with respect to the laminar predictions are some preliminaries of what will come for significantly lower Pr or higher Ra.

To summarize, we generalized the dynamical rescaling method of Zhou and Xia \cite{zho10,zho10b,zho11} for the first time to the data obtained from fully resolved 3D DNS of RB convection in a cylinder. In agreement with the results obtained in 2D we find that the method clearly reveals the temperature profile of a laminar, zero-pressure gradient BL according to Pohlhausen in 3D flows. The method works best for relative low $Ra$ and high $Pr$, where a relatively strong LSC is formed. For lower $Pr$ and higher $Ra$ the velocity field  becomes more turbulent and advects the temperature differently than described by purely laminar theory. Therefore, the agreement between the Pohlhausen profile and the dynamically rescaled profile becomes less for lower $Pr$, which can be explained by the fact that the viscous BL is thicker than the thermal BL for high $Pr$ and hence the thermal BL in this case is less influenced by the bulk turbulent flow than at low $Pr$. For high $Ra$ the effect of passing plumes leads to a larger deviations with respect to the Pohlhausen temperature profile.

We gratefully acknowledge the support from the Foundation for Fundamental Research on Matter (FOM) and the National Computing Facilities (NCF), both sponsored by NWO (R.J.A.M.S. and D.L.), from the Natural Science Foundation of China (No. 11002085) and ``Pu Jiang" project of Shanghai (No. 10PJ1404000) (Q.Z.), and from the Research Grants Council of Hong Kong SAR (No. CUHK404409) (K.Q.X.). The computations in this project have been performed on the Huygens supercomputer of SARA in Amsterdam, and the HLRS cluster in Stuttgart as part of a large scale computing project.

\end{document}